cond-mat/9905175 12 May 1999

# Application of Raman Scattering to Study the Strain Distribution in SiGe Layers

Zhongwei Pan*
*Department of Electrical Engineering, the City College of the City University of New York, New York, NY 10031*

Richard Chin
*Computer Network Services Inc., 100 Ford Road, Denville, NJ 07834*

## Abstract

In this paper, we mainly present a new simple, efficient and nondestructive Raman scattering analysis method to study the distribution of strain in SiGe alloy films. The method can simultaneously determine both Ge fractional composition x and strain of SiGe films. We use the Ar+ laser lines (514.5, 488 and 457.9 nm) to detect the information of different depth of SiGe layers. The results show that the x value (about 0.50) of $Si_{1-x}Ge_x$ and the strain of the films decrease from the interface between SiGe layers and Si substrates to SiGe surface.

## INTRODUCTION

Traditionally, Si integrated circuits such as those found in computers, appliances, toys, and many other applications have used Metal Oxide Semiconductor Field Effect Transistors (MOSFETs) and Bipolar Junction Transistors (BJTs), but neither of these transistors operates above a few gigahertz because of the material properties of Si. Recently, a new material system based on epitaxial SiGe layers on Si has been developed which permits the fabrication of Heterostructure Bipolar Transistors (HBTs), Modulation Doped Field Effect Transistors (MODFETs), and infrared detectors that operate into the millimeter wave frequency spectrum and thus make Si a viable material for microwave circuit applications. It is very important to control Ge fraction x and strain which is caused by the lattice mismatch of Si-Ge layer on Si substrates for fabrication of SiGe devices. Conventionally the combination of standard x-ray double-crystal diffractometry (XDCD) and transmission electron microscopy (TEM) measurements can determine Ge content and strain[1], but owing to XDCD's strong penetrative ability, this method cannot obtain further more information about different depth of SiGe layer. In this paper, we present a new Raman scattering measurement method which not only can easily determine x value of Ge and strain of SiGe layers, but also can receive the strain distribution in different depth of SiGe films.

## DEVICE PREPARATIONS

Epitaxial $Si_{1-x}Ge_x$ layers were grown by APCVD on Si (100) substrates in an Applied Materials 1200 vertical epitaxy reactor. The growth temperature was about 1000 °C with a typical deposition rate ~ 1.0 μm/min. Layer thickness varied between 0.1 and 10 μm. APCVD has the advantage that the resulting epilayers have defect densities that are lower than those achieved by other growth techniques[2-4]. Rutherford backscattering spectroscopy (RBS) was used to determine the SiGe layer thickness. The sample device, which was all uncapped, was a single quantum well (SQW). The SiGe thickness was about 100 nm. It had defect densities well below $10^4$ $cm^{-2}$ and an average dislocation spacing of 3.2 μm measured by XTEM.

## EXPERIMENTAL RESULTS AND DISCUSSIONS

The Raman spectrum of $Si_{1-x}Ge_x$ consists of three main peaks which are due to Si-Si (~500$cm^{-1}$),Ge-Ge (~300$cm^{-1}$) and Si-Ge (~400$cm^{-1}$), peak vibrations[5] are sensitive to Ge content and strain of the alloy. In the region between the Si-Ge and the Si-Si peaks, some minor features are also present, and attributed to the long-range order[6]. Figure 1 shows frequency shift tendency (i.e. energy shift) of the three phonon modes vs different Ge composition x without stress throughout the whole composition range. In general, because of the strain caused by lattice mismatch, the position of main peaks would have a blue-shift for compressive strain and a red-shift for tensile strain, respectively.

Cerdeira[7] pointed out that the energy shift of the Raman phonon modes, which is induced by strain, could be obtained from the following equation.

$$\frac{\Delta\Omega}{\omega_0} = -k\varepsilon_{11} \qquad (1)$$

where $\Delta\Omega$ is the frequency shift relative to the stress, $\omega_0$ is the frequency of the phonon in relaxed single crystal, k is a constant factor and $\varepsilon_{11}$ is strain, respectively.

Suppose that $Si_{1-x}Ge_x$ alloys have a common constant factor $k_{SiGe}$, the equation (1) can be simplified as follows.

* Electronic mail: zhwei@ee-mail.engr.ccny.cuny.edu

$$\frac{\Delta\Omega_{Si}}{\omega_{Si0}} = -k_{SiGe}\varepsilon_{11}$$
$$\frac{\Delta\Omega_{Ge}}{\omega_{Ge0}} = -k_{SiGe}\varepsilon_{11} \qquad (2)$$

where $\Delta\Omega_{Si}$ and $\Delta\Omega_{Ge}$ are the frequency shifts for Si-Si mode and Ge-Ge mode when the alloy layer was strained, respectively; $\omega_{Si0}$ and $\omega_{Ge0}$ are the frequencies of Si-Si mode and Ge-Ge mode when the alloy layer was relaxed, respectively.

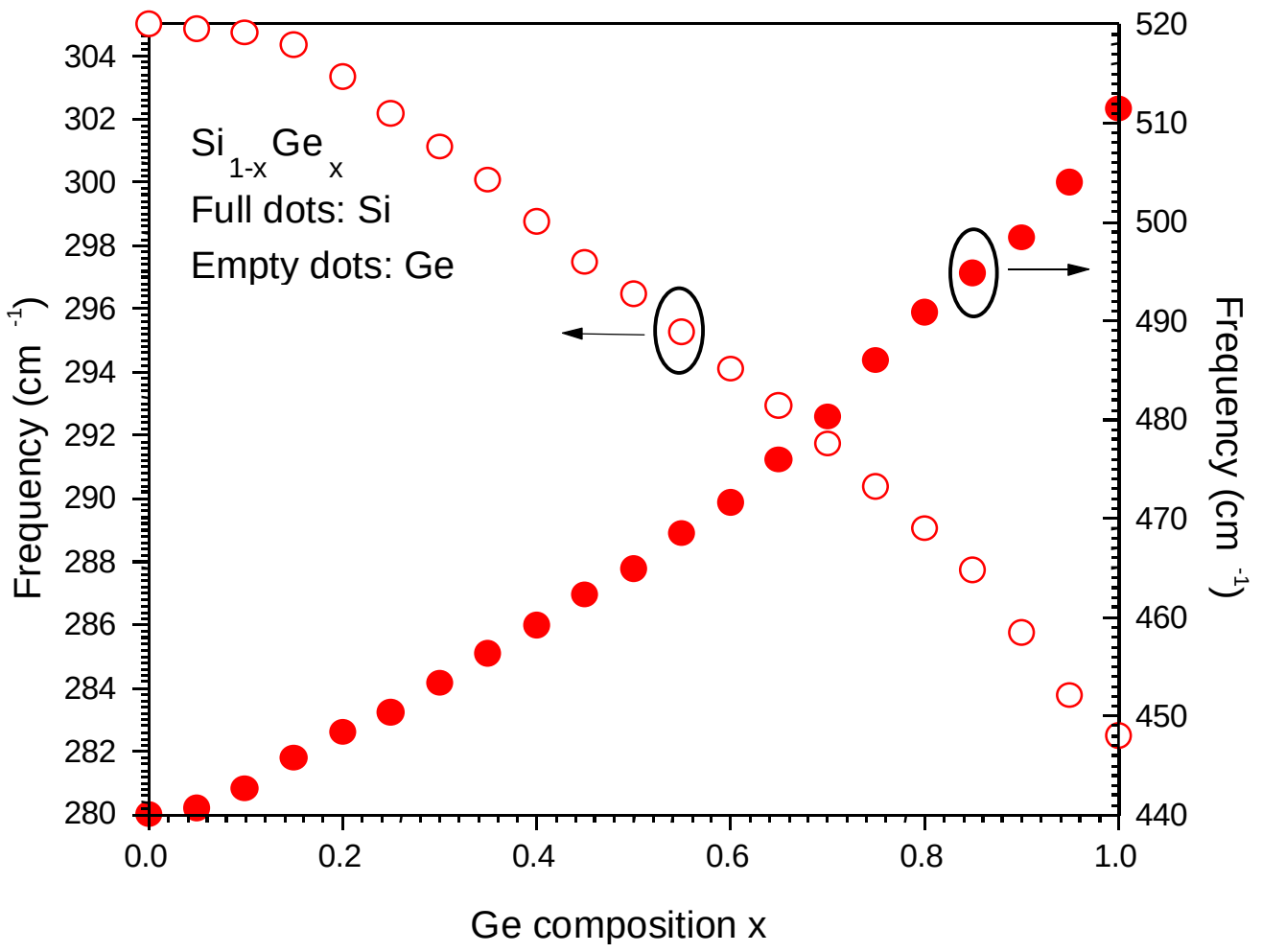

Fig. 1 The relationship between Ge molar composition and the position of $Si_{1-x}Ge_x$ Raman main peaks including Si-Si main peak and Ge-Ge main peak.

Let's assume that $\omega_{Si}$ and $\omega_{Ge}$ are the frequencies of Si-Si mode and Ge-Ge mode when the alloy layer was strained, then we can receive the following equation from equation (2).

$$\omega_{Si} = \omega_{Si0}(1 - k_{SiGe}\varepsilon_{11})$$
$$\omega_{Ge} = \omega_{Ge0}(1 - k_{SiGe}\varepsilon_{11}) \qquad (3)$$

Obviously we can observe the ratio ($\omega_{Si}/\omega_{Ge}$) is a constant when the alloy is strained.

$$\frac{\omega_{Si}}{\omega_{Ge}} = \frac{\omega_{Si0}}{\omega_{Ge0}} \qquad (4)$$

The curve of the ratio vs Ge composition x is shown in Figure 2. Finally we can calculate the strain by the following equation.

$$k_{SiGe}\varepsilon_{11} = 1 - \frac{\omega_{Si} - \omega_{Ge}}{\omega_{Si0} - \omega_{Ge0}} \qquad (5)$$

Room temperature Raman spectra were obtained in a near-backscattering geometry with cross polarization of the incident and scattered light. The $Ar^+$ laser, whose excitation wavelengths were 514.5, 488 and 457.9 nm respectively, was used as exciting source to detect signals from different depth in SiGe alloy layers. When Ge composition x was around 0.5, the Raman penetration depth for the three $Ar^+$ laser lines were 30, 20 and 10 nm respectively (see Figure 3). The 349 and 367 $cm^{-1}$ peaks were identified as 496.5 and 465.7 nm $Ar^+$ laser lines, the 270 and 378 $cm^{-1}$ peaks were Raman resonance peaks, the 289 and 297 $cm^{-1}$ were Ge-Ge LO peaks, the 486 and 498 $cm^{-1}$ were Si-Si LO peaks, the 400 and 405 $cm^{-1}$ were Ge-Si peaks, respectively. The 519$cm^{-1}$ peak came from Si substrate and it was 1 $cm^{-1}$ lower than that of crystalline Si without strain. It shows Si layer of interface has tensile strain. More detailed information is listed in Table 1.

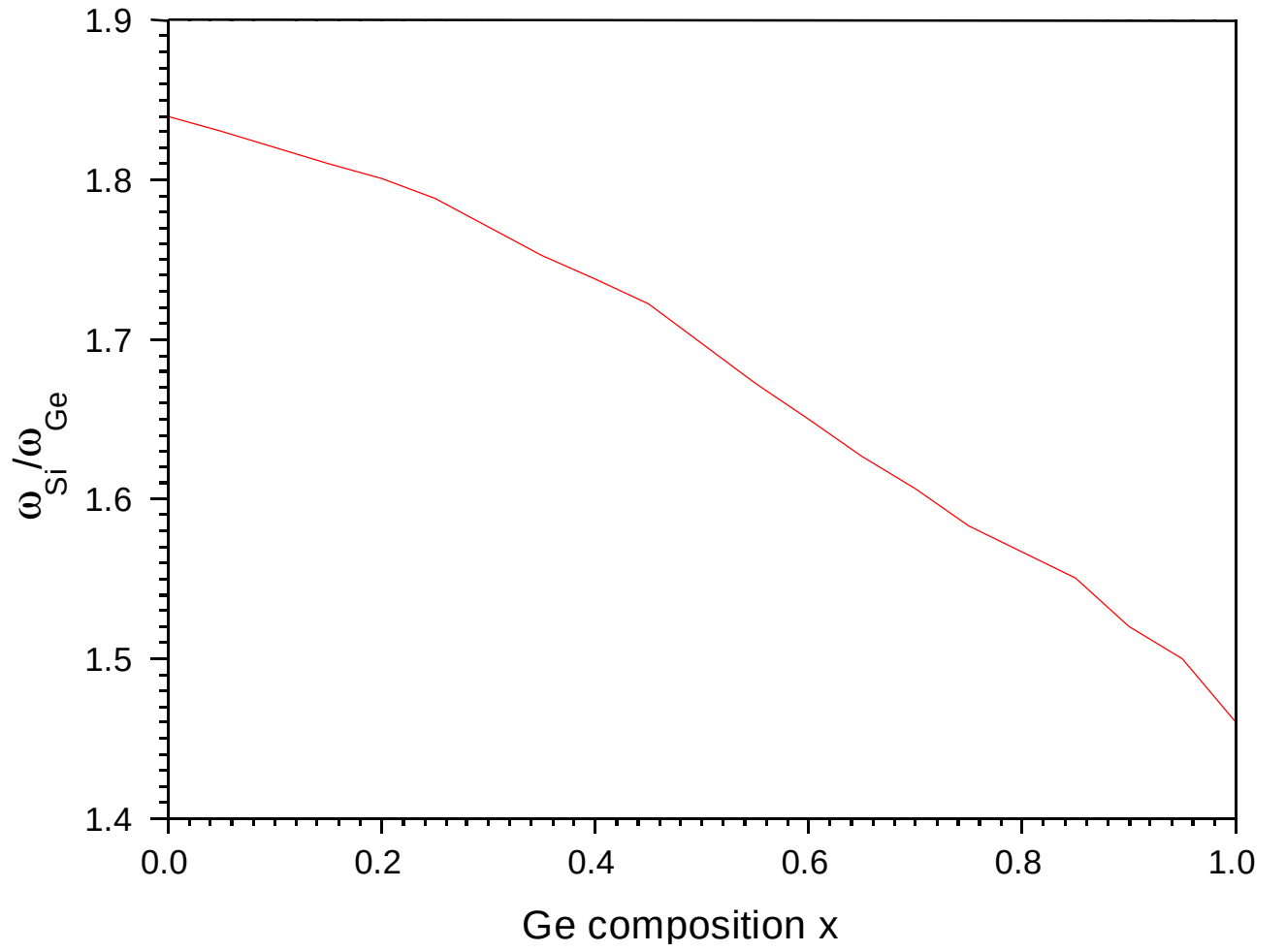

Fig. 2 shows the curve of $\omega_{Si}/\omega_{Ge}$ versus Ge composition x.

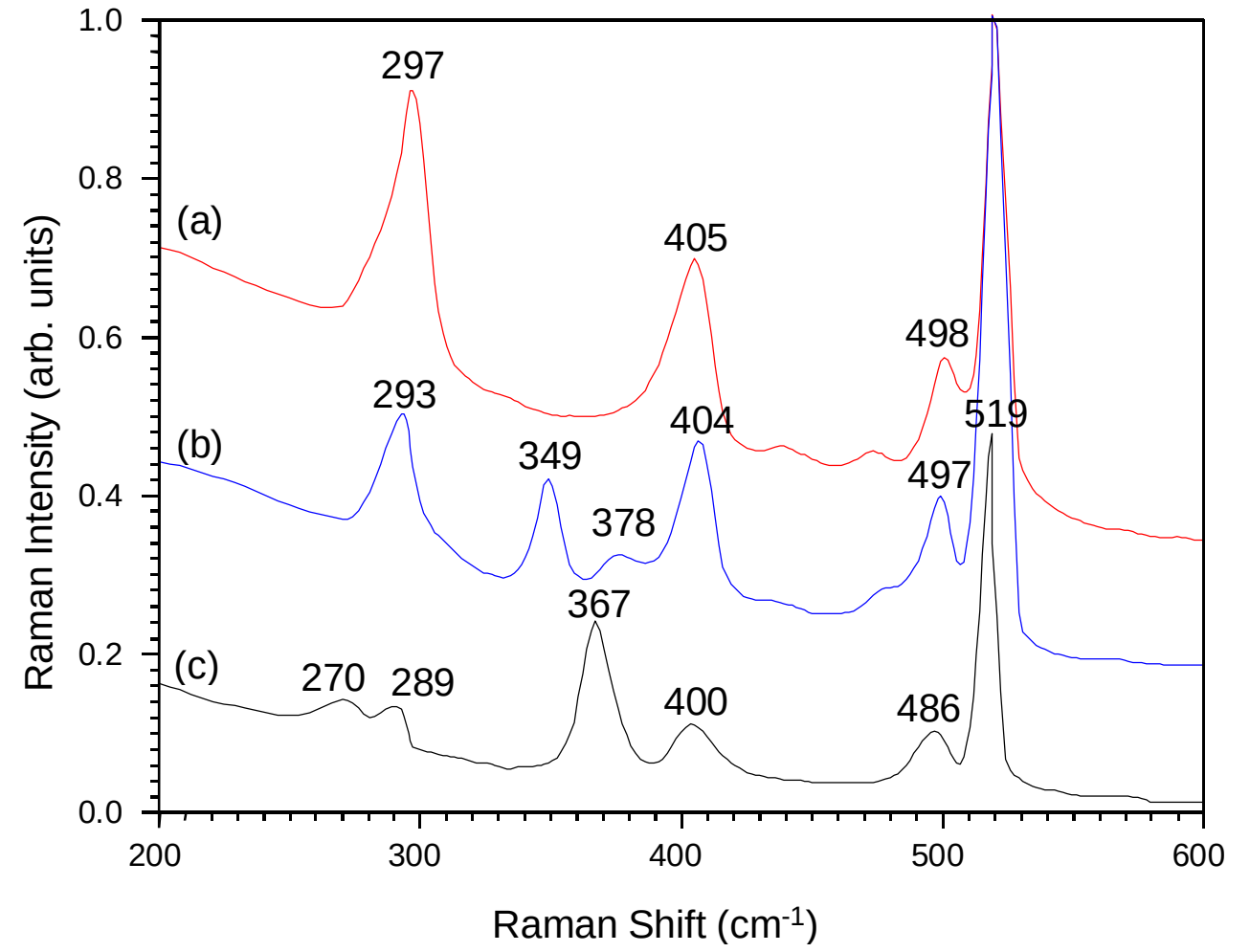

Fig. 3 Raman spectra of SiGe alloy by using different laser lines: (a) 514.5 nm, (b) 488.0 nm and (c) 457.9 nm.

Table 1. The results of Raman scattering

| $Ar^+$ laser line (nm) | Depth (nm) | $\omega_{Si}$ ($cm^{-1}$) | $\omega_{Ge}$ ($cm^{-1}$) | $\omega_{Si}/\omega_{Ge}$ | x | $\varepsilon_{11}$ (%) |
|---|---|---|---|---|---|---|
| 514.5 | 30 | 498 | 297 | 1.68 | 0.53 | -1.93 |
| 488.0 | 20 | 497 | 293 | 1.70 | 0.48 | -1.09 |
| 457.9 | 10 | 496 | 289 | 1.72 | 0.43 | -0.34 |

x: Ge composition

Figure 4 shows that in upper layer, strain relaxed partly attributes to thread dislocations caused by lattice mismatch. The compression strain varies from -0.34% at SiGe surface to -1.93% at interface of SiGe/Si. Because of different diffusion rate (i.e. Si atoms diffuse more speedily in Ge than Ge atoms in Si), the Ge composition x is varied from 0.43 to 0.53 at same region. Further investigation is in progress.

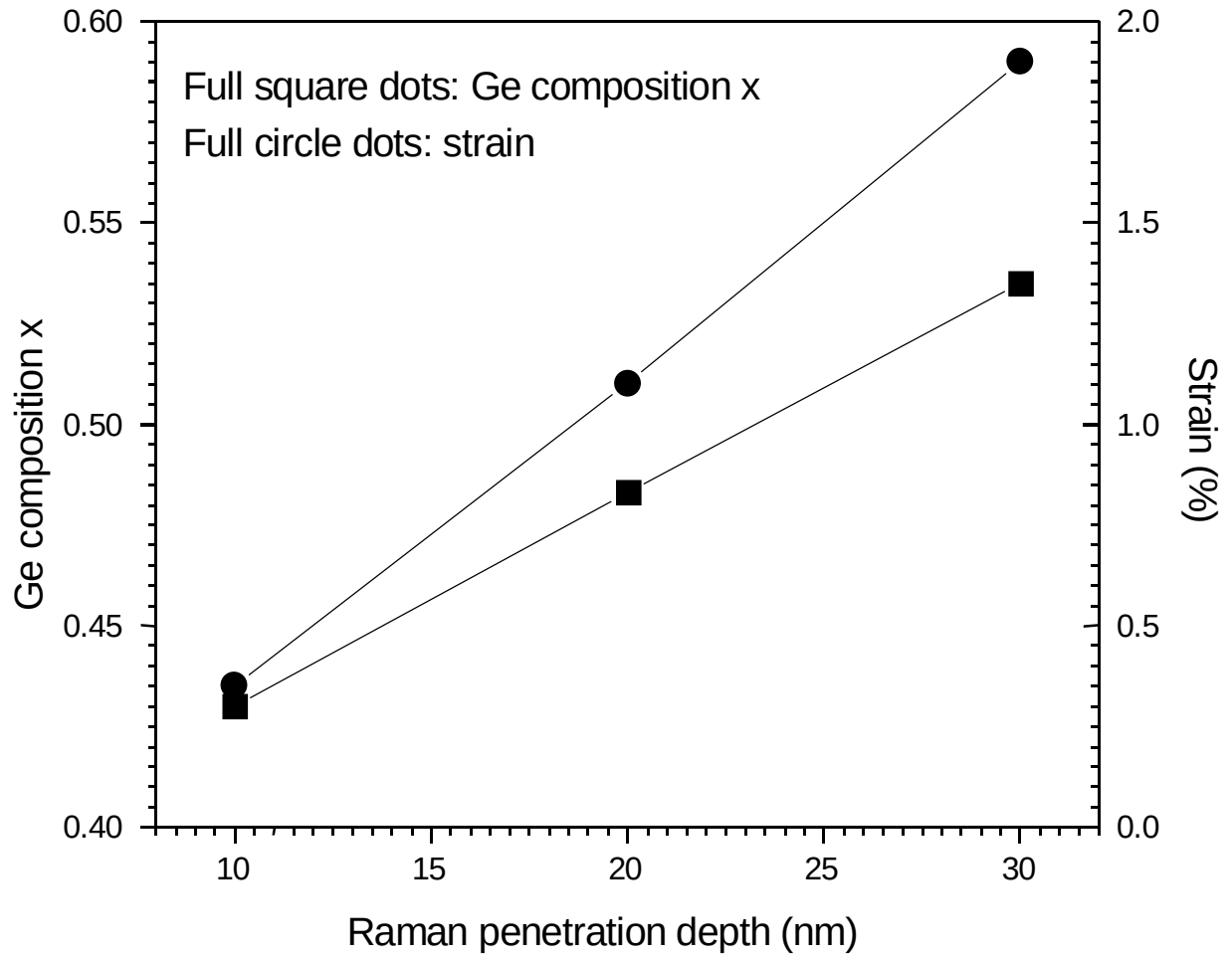

Fig.4 Distribution of Ge composition x and strain in $Si_{1-x}Ge_x$.

## CONCLUSIONS

We present a new method using Raman scattering to investigate the distribution in $Si_{1-x}Ge_x$ alloy layers. During the measurement three $Ar^+$ laser lines are used to detect Raman signals of different depth of SiGe alloy layers. No matter whether the films are relaxed or strained, both experimental and theoretical results are nearly concordant (the ratio of $\omega_{Si}/\omega_{Ge}$ is nearly constant). The analysis technique can conveniently obtain both Ge molar fraction and strain in SiGe alloy layers. The results of measurement show that the strain of upper layers is less than that of interface of alloy and Si substrate. The result indicates that upper layers are partly relaxed. The variation of Ge composition x was caused by diffusion property of Ge and Si. Such measurements offer a convenient, effective and nondestructive method to obtain the information on the structural and compositional features of heteroepitaxial layers.

## ACKNOWLEDGMENTS

The authors would like to thank the support of the New York State Science and Technology Foundation through its CUNY Center for Advanced Technology on Photonic Materials and Applications. Financial support from Reveo Inc. is gratefully acknowledged.

## REFERENCES

[1] E. Bugiel, and P. Zaumseil, Appl. Phys. Lett. **62**, 2051 (1993).
[2] G. L. Patton, J. H. Comfort, B. S. Meyerson, *et. al*, Electron Device Lett. **EDL-11**, 171 (1990).
[3] J. C. Bean, Proc. IEEE **80**, 571 (1992).
[4] N. L. Rowell, J. P. Noel, D. C. Houghton, and M. Buchanan, Appl. Phys. Lett. **48**, 963 (1991).
[5] W. J. Brya, Solid State Commun. **12**, 253 (1973).
[6] D. J. Lockwood, *et. al*, Solid State Commun. **61**, 465 (1987).
[7] F. Cerdeira, C. J. Buchenauer, F. H. Pollak, and M. Cardona, Physical Review B **5**, 580 (1972).